\documentstyle{mn}

\begin{document}

\title[Magnitude of HB stars]
{ THE ABSOLUTE MAGNITUDE OF FIELD METAL-POOR HORIZONTAL BRANCH STARS
\thanks{Based on data from the ESA Hipparcos astrometry satellite.} }

\author[R.G. Gratton] {Raffaele G. Gratton\\
Osservatorio Astronomico di Padova, Vicolo dell'Osservatorio 5, 35122
  Padova, ITALY, gratton@pdmida.pd.astro.it}

\maketitle

\begin{abstract}

HIPPARCOS satellite parallaxes for 22 metal-poor field horizontal branch stars
with $V_0<9$\ are used to derive their absolute magnitude. The weighted mean
value is $M_V=+0.69\pm 0.10$\ for an average metallicity of [Fe/H]=$-1.41$; a
somewhat brighter average magnitude of $M_V=+0.60\pm 0.12$\ for an average
metallicity of [Fe/H]=$-1.51$\ is obtained eliminating HD17072, that might be
on the first ascent of the giant branch rather than on the horizontal branch.
The present values agree with determinations based on proper motions and
application of the Baade-Wesselink method to field RR Lyraes; they are from 0.1
to 0.2~mag fainter than those based on calibration of cluster distances
obtained by using local subdwarfs, and on alternative distance calibrators for
the LMC. The possibility that there is a real difference between the luminosity
of the horizontal branch for clusters and the field is briefly commented. 

\end{abstract}


\newpage

\section{INTRODUCTION}

The determination of the correct distance scale for metal-poor objects has a
large impact on a wide range of astrophysical problems, including the
derivation of ages of globular clusters (a stringent lower limit to the age of
the Universe), of the extragalactic distance scale (affecting the determination
of the Hubble constant), as well as important test on stellar evolution models.
A long-standing lively debate divides the astronomical community amongst
supporters of a "short" and a "long" distance scale: adoption of either of
these two scales would have a deep influence on models for the universe, or for
the formation of our own Galaxy (see e.g. Sandage 1993). 

The recent distribution of the catalogue of calibrated trigonometric parallaxes
measured by the HIPPARCOS satellite (Perryman et al. 1997) has provided new
opportunities for accurate estimates of this distance scale. Various authors
(Reid 1997; Gratton et al. 1997b; Pont et al. 1997) have used parallaxes of
nearby subdwarfs to calibrate the distances to globular clusters. Results
obtained by these three papers are significantly different: this is because
different reddening and metal abundance scales were adopted for subdwarfs, and
different corrections were applied to the original values in order to take into
account for the presence of undetected binaries. Undirect estimates of the
distances to metal-poor objects have been obtained by considering the LMC
distances based on Cepheids, on turn calibrated against nearby objects (Feast
\& Catchpole 1997). 

An alternative way to use HIPPARCOS parallaxes is to consider horizontal branch
(HB) stars, a traditional distance ladder for metal-poor population. Fernley et
al. (1997) tried to measure directly the distances to RR Lyrae variables:
however only the prototype of this important class of pulsating stars is within
300 pc from the Sun, so that its parallax can be measured with some
reliability. 

In this paper we will use HIPPARCOS parallaxes for three RR Lyraes, for a
sample of nine red HB stars, selected on the basis of Str\"omgren photometry
colours, and for ten field blue HB branch stars. Consideration of these other
stars substantially enlarge the sample of nearby metal-poor HB stars. The
sample is presented in Section 2; in Section 3 we discuss the derivation of the
absolute magnitudes; finally the impact of the present results is briefly
discussed in Section 4. 

\section{THE HORIZONTAL BRANCH SAMPLE}

\subsection{The red horizontal branch stars}

The candidates field red horizontal branch (RHB) stars have been identified on
the basis of the reddening corrected $c_1$, $b-y$\ diagram published by
Anthony-Twarog \& Twarog (1994: see their Figure 8). Stars belonging to the red
HB are clearly identified in this diagram as metal-poor stars with
$0.28<(b-y)_0<0.46$\ and $c_0\sim -1.815 (b-y)_0+1.26$. In order to enlarge the
original sample of five stars with $V<9$\ in the list of Anthony-Twarog \&
Twarog, we considered all stars listed in the homogeneized catalogue of
Str\"omgren photometry (Hauck \& Mermilliod 1990) occupying a similar position
in the $c_1$, $b-y$\ diagram (we also considered the catalogues by Olsen 1993,
1994a, 1994b, but they did not provide additional good candidates). The
selection criteria were $V<9$, $0.2<(b-y)<0.5$, $0.4<c_1<0.8$, $m_1<0.15$, and
$-0.1<[c_1+2\,(b-y)-1.2]<0.2$. 86 stars passed these selection criteria: most
of them are highly reddened A and F dwarfs. However, only ten stars have proper
motion $>0.065$~mas/yr from the Hipparcos catalogue (corresponding to a
transverse velocity $>130$~km~s$^{-1}$\ for $V=9$\ and $>80$~km~s$^{-1}$\ for
$V=8$\ HB stars): with the exception of HD165908 (a dwarf at about 16 pc from
the Sun, projected toward a direction very close to that of the galactic
centre), all other high proper motion stars selected by this procedure are very
good candidate RHB stars. Note that all the RHB stars listed by Anthony-Twarog
\& Twarog were recovered by this procedure. Basic data are given in Table~1. 

\begin{table*}
\begin{minipage}{160mm}
\caption{Data for the HB sample}
\begin{tabular}{rrrlllrccrr}
\hline
    HD & $V$ &$B-V$&$b-y$&$m_1$&$c_1$&$E(B-V)$&[Fe/H]&$\pi$&$M_v$&
$\delta M_V$\\
\hline
\\
\multicolumn{11}{c}{Blue Horizontal Branch stars}\\
\\
 31943 &8.254 &0.121 &0.083&0.142&1.226&0.015&         & $3.88\pm 0.74$ &
$ 1.15\pm 0.41$ & $-0.09$ \\ 
 74721 &8.710 &0.042 &0.029&0.127&1.273&0.000& $-1.43$ & $0.34\pm 1.46$ &
$-3.63\pm 9.32$ & $-0.30$ \\
 86986 &8.000 &0.119 &0.092&0.109&1.278&0.035& $-1.88$ & $3.78\pm 0.95$ &
$ 0.78\pm 0.55$ & $-0.13$ \\
 93329 &8.780 &0.129 &0.060&0.123&1.315&0.156& $-1.39$ & $3.89\pm 1.09$ &
$ 1.25\pm 0.61$ & $-0.60$ \\
109995 &7.605 &0.047 &0.050&0.117&1.305&0.001& $-1.78$ & $4.92\pm 0.89$ &
$ 1.06\pm 0.39$ & $-0.28$ \\
128801 &8.739 &$-0.038$&$-0.005$&0.109&1.056&0.037&$-1.17$&$2.40\pm 1.18$&
$ 0.53\pm 1.07$ & $-0.86$ \\
130095 &8.134 &0.032 &0.064&0.108&1.256&0.064& $-1.92$ & $5.91\pm 1.08$ &
$ 1.79\pm 0.40$ & $-0.62$ \\
139961 &8.861 &0.098 &0.077&0.115&1.298&0.107&         & $4.50\pm 1.19$ &
$ 1.80\pm 0.57$ & $-0.52$ \\
161817 &6.978 &0.166 &0.126&0.100&1.197&0.020& $-1.69$ & $5.81\pm 0.65$ &
$ 0.74\pm 0.24$ & $-0.03$ \\
167105 &8.960 &0.020 &0.036&0.120&1.260&0.057& $-1.89$ & $3.02\pm 1.78$ &
$ 1.18\pm 1.28$ & $-0.64$ \\ 
\\
\multicolumn{11}{c}{RR Lyrae variables}\\
\\
RR Lyr &7.66  &      &     &     &     &0.03 & $-1.39$ & $4.38\pm 0.59$ &
$0.77\pm 0.29$  &  0.00  \\
RZ Cep &9.57  &      &     &     &     &0.35 & $-1.59$ & $0.22\pm 1.09$ &
$-4.80\pm 10.75$  &  0.00  \\
MT Tel &9.10  &      &     &     &     &0.045& $-2.00$ & $1.01\pm 1.26$ &
$-1.02\pm 2.71$  &  0.00  \\
\\
\multicolumn{11}{c}{Red Horizontal Branch stars}\\
\\
 17072 &6.610 &0.660 &0.441&0.132&0.452&0.011& $-0.77$ & $7.57\pm 0.51$ &
$0.97\pm 0.15$ & $-0.02$ \\
 18550 &8.260 &0.455 &0.29 &0.12 &0.74 &0.004& $-0.66$ & $3.35\pm 1.06$ &
$0.87\pm 0.69$ & $-0.01$ \\
 25532 &8.208 &0.657 &0.482&0.094&0.507&0.074& $-1.08$ & $4.39\pm 1.25$ &
$1.19\pm 0.62$ & $-0.02$ \\
 97650 &7.880 &0.653 &0.45 &0.14 &0.43 &$-$0.008&$-0.98$&$3.13\pm 0.99$ &
$0.38\pm 0.69$ & $-0.02$ \\  
105546 &8.610 &0.660 &0.460&0.130&0.420&0.000& $-1.33$ & $1.96\pm 0.91$ &
$0.07\pm 1.01$ & $-0.02$ \\
106373 &8.918 &0.442 &0.338&0.031&0.732&0.061& $-2.74$ & $2.22\pm 1.09$ &
$0.46\pm 0.92$ &   0.00  \\
107550 &8.350 &0.719 &0.50 &0.14 &0.43 &0.054& $-0.78$ & $2.81\pm 1.07$ &
$0.43\pm 0.83$ & $-0.02$ \\
184266 &7.600 &0.548 &0.427&0.063&0.605&0.054& $-1.93$ & $3.28\pm 0.95$ &
$0.01\pm 0.63$ & $-0.01$ \\
208360 &7.630 &0.640 &0.46 &0.12 &0.45 &0.018& $-1.17$ & $4.71\pm 0.93$ &
$0.94\pm 0.43$ & $-0.02$ \\
\hline
\end{tabular}
\end{minipage}
\end{table*}

The interstellar reddening and metallicity were obtained following the same
procedure adopted by Anthony-Twarog \& Twarog (1994, ATT). In order to compare
the results obtained in this paper with other estimates for the absolute
magnitude of the HB, we put the metal abundances listed by Anthony-Twarog \&
Twarog on the same system as those considered in Gratton et al. (1997b). To
this purpose, we considered all stars in the original sample of Anthony-Twarog
\& Twarog which have metallicities measured by Gratton, Carretta \& Castelli
(1997a, GCC). We found that the mean relation between the two values of
metallicity is: 
\begin{equation}
{\rm [Fe/H]}_{\rm GCC}=(1.33\pm 0.23) {\rm [Fe/H]}_{\rm ATT} +
(0.56\pm 0.32). 
\end{equation}
The values listed in Table 1 have been obtained from the original values listed
by Anthony-Twarog \& Twarog after application of eq. (1). 

\subsection{RR Lyrae}

The sample of RR Lyrae was taken from Preston, Schectman \& Beers (1991): it
includes the only three known RR Lyrae variables with an average dereddened 
$V_0$\ magnitude brighter than 9. Metallicity and reddening for RR Lyr were
taken from Clementini et al. (1995); those for RZ Cep and MT Tel were taken
from Preston et al. (the $\Delta S$\ calibration of Clementini et al. was
used). For RR Lyrae we assumed an average $V$\ magnitude of $V=7.66\pm 0.05$\
(Layden 1994); the rather large uncertainty is due to the Blazhko effect that
is important for this star (note however that this uncertainty is much smaller
than the error bar due to the parallax). For the two other stars we used the
$V_0$\ values of Preston et al.

\subsection{The blue horizontal branch stars}

The blue horizontal branch (BHB) stars considered in this paper are taken from
the list of Stetson (1991). Again, only stars with $V<9$\ were considered. We
found 12 such stars in the list of Stetson; parallax is not available for one
of them (HD203653). According to Hipparcos, HD214539 has a moderate proper
motion of $\mu=0.036$~arcsec/yr, all other metal-poor HB stars considered in
this paper having $\mu>0.065$~arcsec/yr; it also has a small value of the
Str\"omgren $c_1$\ index of $c_1=1.036$, not compatible with its positive
$b-y$\ colour if this star is on the HB. These two facts suggest that this star
is not a BHB star; we finally decide to leave this star out from our sample. 

We tried to enlarge this sample by considering all stars in the catalogue of
homogeneized Str\"omgren photometry (Hauck \& Mermilliod 1990) having colors
similar to those of known BHB; however no additional star with proper motion
larger than $\mu=0.05$~arcsec/yr was found. We then finally considered ten BHB
stars. 

Main parameters are listed in Table 1. $V$\ magnitudes are the average of those
listed in the Hipparcos catalogue and those by Stetson (1991); $B-V$\ colours
were taken from the Hipparcos catalogue, while Str\"omgren photometry was from
Stetson (1991). Whenever possible, reddenings are those listed by Gray,
Corbally \& Philip (1996). However, this was possible for only five of the
programme stars. For the remaining objects, reddening was obtained by comparing
the $\beta$\ index (from Stetson) with the apparent $B-V$\ colour, assuming
that the following relation (obtained by considering stars with reddening from
Gray et al.) holds for BHB stars: 
\begin{equation}
\beta = 2.861 + 0.188 (B-V)_0 - 6.581 (B-V)_0^2.
\end{equation}
For some objects there is ambiguity on the reddening determined by this
procedure, two values being compatible with data. We initially assumed the
lowest reddening solution for all stars; however, we found that the higher
reddening solution is required to explain the observations for HD93329 and
HD139961. Our reddening estimates are very close to the values adopted by De
Boer, Tucholke, \& Schmidt (1997): the average difference for the five stars in
common is $0.005\pm 0.008$~mag (single star standard deviation of 0.018~mag),
in the sense that our reddening are smaller. 

Metallicities were taken from Adelman \& Philip (1996). The metallicity scale
for the BHB stars is quite uncertain, due to the possible effects of diffusion
in the subatmospheric layers, and of departures from LTE in the photospheric
regions. However, these effects are ignored in our discussion. For the two
stars missing metal abundance values (HD31943 and HD139961), we assumed a metal
abundance equal to the average provided by the other BHB stars
([Fe/H]=$-1.66$). 

\subsection{Sample uncompleteness}

Our total sample consists of 20 field horizontal branch stars with $V\leq 9$\
(8 with $V\leq 8$, 2 with $V\leq 7$), and two slightly fainter stars. We
roughly estimate uncompleteness of our sample by comparing these numbers with
the list of metal poor dwarfs ([Fe/H]$<-0.9$) compiled by Gratton et al.
(1997c); this list provides an extensive though still uncomplete census of
metal poor dwarfs with $V<10.3$\ and $5<M_V<6.5$. Within this sample of
subdwarfs, we have 12 stars with $5.0<M_V<5.5$\ (closer than $\sim 100$~pc from
the Sun), 9 stars with $5.5<M_V<6.0$\ (closer than $\sim 80$~pc), and 9 stars 
with $6.0<M_V<6.5$\ (closer than $\sim 60$~pc). Assuming completeness down to
$V=10.3$\ (surely an upper limit), this yields a total of $\sim 9,000$\ main
sequence stars with [Fe/H]$<-0.9$ and $5<M_V<6.5$\ within 500 pc from the Sun
(where we assumed a uniform distribution and neglected interstellar
absorption). The ratio between HB stars and main sequence stars in this
luminosity range depends on the adopted slope of the initial mass function; it
is $\sim 120$\ for a Salpeter value. Ignoring interstellar absorption and
assuming a typical magnitude of $V\sim 0.6$\ for HB stars, we then expect to
observe $\sim 75$\ metal-poor HB stars ($\sim 20$\ with $V<8$, and $\sim 5$\
with $V<7$) within the limiting magnitude here considered ($V<9$, $\sim
500$~pc). It must be noted that these numbers depend on our assumptions: a
larger number of missing stars is expected due to uncompleteness of the
subdwarf sample here considered; on the other side, galactic distribution of
stars and interstellar reddening should reduce the expected number of
metal-poor HB stars with $V<9$. However, we will assume they represent a rough
estimate of the real number of HB stars; they suggest that our sample includes
about half of the HB stars with $V<8$, and roughly 20\% of those with $8<V<9$.
Missing stars are more likely to be metal-rich (half of the present sample is
made of stars with [Fe/H]$<-1.5$, while they are only a third of the comparison
subdwarf sample) and/or to have moderate or low proper motion ($\sim 1/3$\ of
the subdwarfs of the comparison sample have transverse velocities
$<100$~km~s$^{-1}$). Str\"omgren photometry may also be missing for a
considerable fraction of the field HB stars with $V>8$. 

\section{MEAN ABSOLUTE MAGNITUDES}

A basic problem in the derivation of mean absolute magnitudes when errors on
parallaxes are similar to the measured values is to give the appropriate weight
to individual stars. The usual procedure is to average absolute magnitude
values with weights given according to the individual errors in the absolute
magnitude: this is simply $2.17~\Delta\pi/\pi$, where $\pi$\ is the parallax
and $\Delta\pi$\ its error. This procedure overestimates weights given to those
objects having parallaxes measured too high (that is located closer than they
really are) with respect to those objects having parallaxes measured too low
(that is located farther than they really are). The net effect is that
distances are underestimated, and hence on average the mean absolute magnitude
is estimated to be fainter than really is. This is one of the causes of the
so-called Lutz-Kelker effect (Lutz \& Kelker 1973). 

In order to give same weight to parallaxes measured too high and too low, we
decided to average the parallaxes rather than the absolute magnitudes. In
practice, we assumed that all stars have the same absolute magnitude $M$, apart
from a term dependent on the colour $\delta M(B-V)$\ which accounts for the not
perfect horizontality of the HB; this value was determined by interpolation on
the mean loci of M5 (Sandquist et al. 1996), for which we assumed a reddening
of $E(B-V)=0.035$\ (Gratton et al. 1997b) and a magnitude of $V=15.11$\ at
centre of the RR Lyrae strip. M5 was selected because it has an extended HB and
accurately calibrated photometry available. It should however be noted that M5
([Fe/H]=$-1.17$: Sneden et al., 1992; [Fe/H]=$-1.11$: Carretta \& Gratton,
1997) is slightly more metal-rich than the bulk of the program HB stars
([Fe/H]$\sim -1.5$). The expected value of the parallax $\pi^*$\ corresponding
to this value of $M$\ is simply given by: 
\begin{equation}
\pi^*=10^{0.2\,[M-V-\delta M(B-V)]-1+0.62\,E(B-V)},
\end{equation}
where $V$ is the apparent magnitude and $E(B-V)$\ is the interstellar reddening.
The value of $M$\ is then derived by assuming that the weighted average of
$\pi-\pi^*$\ is zero.

Two other systematic corrections should in principle be applied to our sample:
\begin{itemize}
\item a fraction of the programme stars is expected to have a companion. The
total magnitude of the system is then expected to be brighter than for single
stars. This correction should be small, because it is unlikely that the
companion is also an evolved star. To estimate this correction, we computed the
average correction due to a secondary which has the same luminosity function
(transformed from $B$\ to $V$\ magnitudes) of M5 observed by Sandquist et al.
(1996). The average correction for a binary system made of an HB star and a
secondary brighter than $M_V\sim 7$\ is estimated to be 0.019 mag. Only a
fraction of the observed HB stars have such a bright companion. We then
estimate that the binary correction for the whole sample is $<0.01$~mag. We
will neglect this small correction
\item star selection has a threshold in the apparent $V$\ magnitude ($V<9$). A
Malmquist bias is then expected to be present, favouring intrinsically bright
objects. The correction is expected to be quite small, because HB stars have a
small intrinsic scatter in their absolute magnitudes (at a given metallicity,
peak-to-peak dispersion is $\leq 0.3$~mag). We estimated this correction by
means of Monte Carlo simulations, assuming a uniform space distribution for HB
stars (likely a good approximation for nearby objects). We found that Malmquist
correction is $\sim 0.01$~mag for a uniform distribution of $M_V$'s within a
range of 0.3~mag, and $\sim 0.02$~mag for a range of 0.4~mag (these results
depend only very marginally on the adopted average $M_V$). The actual
distribution is likely to be more peaked than a uniform distribution, so that
these estimates of the Malmquist bias are likely to be somewhat overestimated.
Note also that this Malmquist bias is partly compensated by the selection
criterion based on proper motion, favouring nearby intrinsically faint objects
with respect to far brighter ones; this effect is however expected to be
smaller than the Malmquist bias because it depends linearly rather than
quadratically on distance. This is confirmed by Monte Carlo simulations which
takes into account the kynematical properties of the metal-poor stars: we found
that the correction for the proper motion threshold amounts to less than 0.005
mag for any reasonable value of the parameters. Hereinafter we will neglect
these corrections, being much smaller than other sources of error 
\end{itemize}

Using this procedure, the weighted average absolute magnitude of the HB at the
RR Lyrae colour is: 
\begin{equation}
M_V(RR) = 0.69\pm 0.10.
\end{equation}
The weighted mean metallicity of the stars used to derive this relation is
[Fe/H]=$-1.41$. 

It should be noted that $\sim 30$\% of the total weight in the present
derivation of the absolute magnitude of the metal-poor HB stars is due to a
single star, namely HD17072, for which we found $M_V=0.97\pm 0.15$. This is a
rather metal-rich star ([Fe/H]=$-0.77\pm 0.3$): misclassification of red giant
branch stars as red HB stars is possible when the $c_1$, $b-y$\ diagram is used
for metal-rich objects. It is then possible that this star is on the red giant
branch rather than on the HB. If this star is eliminated, the weighted average
absolute magnitude would be: 
\begin{equation}
M_V(RR)=0.60\pm 0.12
\end{equation}
for an average metallicity of [Fe/H]=$-1.52$. Given its moderately high metal
abundance, the absolute magnitude of HD17072 is marginally consistent with the
value estimated from the remaining stars, if a slope of $\sim 0.2$\ is adopted
in the relation between absolute magnitude and metallicity for HB stars (see
e.g. the discussion in Gratton et al. 1997b). 

If we consider separately RHB, RR Lyrae variables and BHB stars, we found
average magnitudes of: 
\begin{equation}
M_V(RR)=0.75\pm 0.14~~~~(9~{\rm RHB~stars}),
\end{equation}
\begin{equation}
M_V(RR)=0.33\pm 0.31~~~~(3~{\rm RR~Lyrae~variables}),
\end{equation}
 and:
\begin{equation}
M_V(RR)=0.73\pm 0.16~~~~(10~{\rm BHB~stars})
\end{equation}
for average metal abundances of [Fe/H]=$-1.15$, [Fe/H]=$-1.52$, and
[Fe/H]=$-1.66$\ respectively. The result for BHB stars is quite heavily
influenced by the colour correction for the non-horizontality of the HB, which
may be as large as 0.6 mag and it is quite uncertain. However, the average
magnitude changes by only 0.02~mag (up to $M_V(RR)=0.71\pm 0.19$) when only
stars with $(B-V)_0>0$\ are considered. We conclude that our result is only
marginally dependent on the colour correction for the non-horizontality of the
HB. 

\begin{table}
\caption{Average magnitude for field HB stars}
\begin{tabular}{cccc}
\hline
 bin & No. stars & ${\rm [Fe/H]}$ & $M_V(RR)$ \\
\hline
 ${\rm [Fe/H]}<-1.5$        & 11 & $-1.84$ & $0.63\pm 0.17$ \\
 $-1.5<{\rm [Fe/H]}<-1.0$   &  7 & $-1.31$ & $0.58\pm 0.21$ \\
 ${\rm [Fe/H]}<-1.0$        &  4 & $-0.79$ & $0.85\pm 0.15$ \\
\hline
\end{tabular}
\end{table}

The uncertainties present in this estimate for the average magnitude of the HB
preclude an accurate estimate of the dependence of HB magnitude on metallicity.
In fact, if we divide our sample into three groups according to metal abundance
([Fe/H]$<-1.5$, $-1.5<$[Fe/H]$<-1$, and [Fe/H]$>-1$), we find the average
magnitudes listed in Table~2; a weighted mean square fit through these three
points then gives: 
\begin{equation}
M_V(RR)=(0.22\pm 0.22)({\rm [Fe/H]}+1.5)+(0.66\pm 0.11).
\end{equation}
The large error bar for the slope of this relation makes it compatible with all
literature estimates.

\section{DISCUSSION AND CONCLUSIONS}

On the whole, estimates of the absolute magnitude of the HB for metal-poor
stars provide contradictory results. Other determinations from direct
observation of field HB stars lead to values very close to that derived in
this paper. Layden et al. (1996) obtained $M_V(RR)=+0.71\pm 0.12$\ for
[Fe/H]=$-1.6$, and $M_V(RR)=+0.79\pm 0.30$\ for [Fe/H]=$-0.76$\ from
(ground-based) statistical parallaxes of RR Lyrae stars. Similar faint
magnitudes are obtained from application of the Baade-Wesselink method to RR
Lyraes; for instance Clementini et al. (1995) obtained: 
\begin{equation}
M_V(RR)=(0.19\pm 0.03)({\rm [Fe/H]}+1.5)+(0.68\pm 0.04).
\end{equation}
Very similar relations were previously obtained by Carney, Storm \& Jones
(1992) and Fernley (1993); note however that brighter absolute magnitudes and a
steeper slope has been recently obtained by McNamara (1997) from a reanalysis
of Baade-Wesselink results using a higher temperature scale: 
\begin{equation}
M_V(RR)=(0.29\pm 0.05)({\rm [Fe/H]}+1.5)+(0.53\pm 0.05).
\end{equation}
The value for the slope of the [Fe/H]$-M_V(RR)$\ relation is lively debated
(for a discussion, see e.g. Carney et al. 1992). As mentioned above, this slope
cannot be determined from data of the present paper alone. However, if we
arbitrarily adopt the slope given by Clementini et al. (1995), we found: 
\begin{equation}
M_V(RR)=0.19 ({\rm [Fe/H]}+1.5)+(0.66\pm 0.10).
\end{equation}
If we now consider the constant term, the agreement with the value from
Clementini et al. (1995) is fully satisfactory (a marginal agreement is found
also with the relation of McNamara). On the other side, if HD17072 is
eliminated, the relation would be: 
\begin{equation}
M_V(RR)=0.19 ({\rm [Fe/H]}+1.5)+(0.61\pm 0.12),
\end{equation}
still in agreement with the value obtained using the Baade-Wesselink technique.

On the other hand, much brighter magnitudes are obtained by those estimates
based on the HB of globular clusters. As an example, Gratton et al. (1997b)
determined the magnitude of the HB from a calibration of globular cluster
distances based on subdwarfs:
\begin{equation}
M_V(RR)=(0.22\pm 0.09)({\rm [Fe/H]}+1.5)+(0.43\pm 0.04),
\end{equation}
where we consider eq. (11) of Gratton et al., that is more appropriate for a
comparison with field HB stars because it takes into account the evolution of
stars off the zero age horizontal branch (ZAHB). Very recently, Gratton et al.
(1997c) revised this relation to the following: 
\begin{equation}
M_V(RR)=(0.18\pm 0.09)({\rm [Fe/H]}+1.5)+(0.47\pm 0.04),
\end{equation}
using a more extended sample which includes nearly 60 metal-poor subdwarfs with
accurate parallaxes. The average magnitude of the HB obtained in the present
paper is 0.19~mag fainter than the relation provided by the calibration of
globular cluster distances based on subdwarfs. The disagreement is even worse
if the distance scale found by Reid (1997) is considered; and it decreases only
marginally if the distance scale considered by Pont et al. (1997) is adopted.
Note however that the average magnitude we obtain eliminating HD17072 is in
marginal agreement with that found using the calibration of globular cluster
distances based on subdwarfs. 

Another argument in favour of a {\it long} distance scale (longer than found
here) is provided by consideration of the RR Lyrae in various clusters in the
Large Magellanic Cloud (LMC) (with an average [Fe/H]$=-1.9$), whose distance is
assumed to be identical to that of the LMC obtained from other calibrators.
The first determination by Walker (1992) was $M_V=0.44$, for an LMC true
distance modulus of $(M-m)_0=18.50$\ based on pre-Hipparcos calibration of the
period-luminosity relation for the Cepheids. The value has been recently
revised to $M_V=0.24\pm 0.10$\ by Feast \& Catchpole (1997) using a calibration
of the Cepheid period-luminosity relation based on Hipparcos data. Less
extreme, but still bright magnitudes are obtained using the LMC distance
modulus from Hipparcos calibration for Miras ($M_V=0.40\pm 0.2$: van Leeuwen et
al. 1997), and from the expanding ring around SN1987a ($M_V=0.36\pm 0.03$:
Panagia et al. 1997; note however that a fainter value of $M_V>0.50$\ has been
obtained by Gould \& Uza 1997). Even considering a $\sim 0.08$~mag correction
for the metallicity of the LMC clusters, these estimates are brighter than the
present value for the constant term in eqts. (12) and (13), while they agree
with the determination based on local subdwarfs and globular clusters (eqt.
15). 

While a marginal agreement between the distance scale derived from parallax of
field HB stars and those which use globular clusters can be found excluding
HD17072, on the whole it seems that the distance scales provided by field stars
are uncomfortably shorter than those provided by globular clusters. Is there
something wrong in some of these distance derivations? An accurate revision of
each method is beyond the scope of the present discussion; we only note that
the present derivation is independent of stellar models, and it depends only
marginally on the adopted reddening and abundance scales; it is however
sensitive to the attribution of individual stars to the HB: this may be
questioned for some important object. Furthermore, the error bar is quite
large. On the other side, the calibration of the cluster distances using local
subdwarfs may be critically affected by systematic errors in the metal
abundance and reddening scales. For instance, it would be increased by $\sim
0.08$~mag if the metallicity scale for globular clusters (derived from high
dispersion spectroscopic analysis of giants) is systematically too large with
respect to the scale adopted for subdwarfs (derived from analysis of dwarfs)
by 0.1 dex. In both cases the real error bars are then larger than the nominal
one, given by dispersion of individual data. 

It might also be noted that there seems to be good agreement between different
estimates for field HB stars; the same can be said for those estimates using
globular cluster HB stars. This fact might suggest that there is perhaps a real
difference between the magnitude of the HB between globular clusters and the
general field, at the level of 0.1 or 0.2 mag. Theoretical arguments supporting
such a difference have been recently advanced by Sweigart (1997), who 
considered the possibility that the Na-O and Mg-Al anticorrelations observed in
globular cluster red giants, but not in field stars (Kraft 1994, and 
references therein), are due to mixing presumably related to internal rotation.
In this scenario helium is also getting mixed, producing a markedly bluer and 
somewhat brighter HB morphology.

This possibility can be directly tested using RR Lyraes in the LMC. There are
two independent estimates of the average magnitude for field RR Lyrae variables
in the LMC: (i) Kinman et al. (1991) found an unreddened average $<B_0>$\
magnitude of $<B_0>=19.37\pm 0.06$; if we assume an average $<(B-V)_0>$\ colour
of $<(B-V)_0>=0.31$\ (Blanco 1992), the average unreddened $<V_0>$\ magnitude
is $<V_0>=19.06\pm 0.06$. (ii) The average $V$\ magnitude of LMC RR Lyrae
observed by the MACHO project (Alcock et al. 1996) is $<V>=19.4$; if we assume
a value of $E(B-V)=0.10$\ for the LMC bar (Bessell 1991), we obtain an average
unreddened $<V_0>$\ magnitude of $<V_0>=19.09\pm 0.06$, in good agreement with
the result by Kinman et al. These values for the field RR Lyraes in the LMC
are to be compared with an average unreddened $<V_0>$\ magnitude of
$<V_0>=18.94\pm 0.04$\ for cluster variables (Walker 1992). Note that Alcock et
al. gives an average metallicity of [Fe/H]=$-1.7\pm 0.2$\ for the field LMC RR
Lyrae observed by the MACHO project, quite close to the average value of
[Fe/H]=$-1.9$\ given by Walker (1992) for the LMC clusters. On the whole, RR
Lyraes in the LMC field seem then fainter than cluster ones by $0.14\pm 0.08$,
approximately the difference required in the present context. However the
difference might be smaller than given by this comparison. In fact, rather
large uncertainties exist in the reddening estimates for the LMC, the adopted
values depending on the adopted field and on the authors (see Bessell 1991 for
a discussion of this point). If we limit ourselves to the variables discovered
in the fields around NGC 1466, NGC 1841, NGC 2210, NGC 2257, and Reticulum
(summarized by Kinman et al.), and adopt for each field the same reddening
considered for the clusters (Walker 1992), field variables are fainter than
cluster ones by only $0.05\pm 0.02$~mag on average (we give here same weight to
all fields, irrespective of the number of variables discovered, to account for
the possibility that individual clusters are at a distance from us different
than the average field variables). Such a small difference might be entirely
attributed to a small difference in the average metallicity (field stars being
slightly more metal rich than the clusters).

Alternative tests are provided by the HB of M31, and from data on the
Baade-Wesselink method. As to M31, the apparent magnitude of the HB for the
globular clusters are in the range $25.29<V<25.66$\ (Fusi Pecci et al. 1996),
close to the values of $V=25.35$\ and $V=25.45$\ obtained by Rich, Mighell \&
Neill (1996) respectively for the BHB and RHB in the field of G219: from these
values the HB of the field does not seem fainter than the HB's of the clusters.
However the value given by Fusi Pecci et al. (1996) for G219 is $V=25.29$, this
is $\sim 0.1$~mag brighter than the value of Rich et al. for field stars in the
same cluster neighborhood. Of course, it should be reminded that reddening and
metallicity for both clusters and field stars are uncertain. It may also be
noted that the apparent magnitudes of Rich et al. for the BHB and RHB of M31,
coupled with the absolute magnitudes of eqt. (6) and (8), yield estimates of
the absolute distance modulus of $(m-M)_0=24.43\pm 0.16$\ and $24.51\pm 0.14$\
for an interstellar reddening of $E(B-V)=0.06$\ toward G219 (Fusi Pecci et al.
1996). These values agree well with the value of $(m-M)_0=24.43$\ given by
Freedman \& Madore (1990) from the pre-Hipparcos Cepheids calibration; they are
smaller than the value of $(m-M)_0=24.77$\ obtained by Feast \& Catchpole
(1997) using the Hipparcos Cepheid calibration. 

As to the Baade-Wesselink method, the absolute magnitudes of RR Lyrae in M5 and
M92 (Storm et al. 1994) are indeed about 0.2 mag above the mean relation
obtained for field stars (see Fig. 16 of Clementini et al. 1995); however some
of these stars are thought to be evolved objects well above the zero age
horizontal branch; on the other side, the variables in M4 (Liu \& Janes 1990)
fit well the relation for field stars. 

On the whole we regard evidences in favour or against a systematic difference
between the luminosity of field and cluster HB as controversial; more work is
required to settle this point. 

We conclude by noting that the adoption of a short distance scale such as that
suggested by the present data would have an important impact in various
astrophysical problems: e.g. the age of globular clusters would be $\sim
16$~Gyr, incompatible with an Einstein-de Sitter model for the Universe, unless
$H_0<40$~km~s~Mpc$^{-1}$. Further work is required in order to determine the
absolute magnitude of HB stars. 

{}


%
%


\begin{thebibliography}{}

\bibitem[]{} Adelman, S. J., \& Philip, A. G. D. 1996, MNRAS, 280, 285
\bibitem[]{} Alcock, C., et al. 1996, AJ, 111, 1146
\bibitem[]{} Anthony-Twarog, B. J., \& Twarog, B. A. 1994, AJ, 107, 1577
\bibitem[]{} Blanco, V. 1992, AJ, 104, 734
\bibitem[]{} Bessell, M. S. 1991, A\&A, 242, L17
\bibitem[]{} Carney, B.W., Storm, J., \& Jones, R.V. 1992, ApJ, 386, 663
\bibitem[]{} Carretta, E., Gratton, R.G. 1997, A\&AS, 121, 95
\bibitem[]{} Clementini, G., Carretta, E., Gratton, R. G., Merighi, R., Mould,
J. R., \& McCarthy, J. K. 1995, AJ, 110, 2319 
\bibitem[]{} De Boer, K. S., Tucholke, H. -J., \& Schmidt, J. H. K. 1997, A\&A,
317, L23 
\bibitem[]{} Feast, M. W., \& Catchpole, R.M. 1997, MNRAS, 286, L1
\bibitem[]{} Fernley, J.A. 1993, A\&A, 268, 591
\bibitem[]{} Fernley, J., Barnes, T. G., Skillen, I., Hawley, S. L., Hanley,
C. J., Evans, D., Solano, E., \& Garrido, R. 1997, in Hipparcos Venice '97, 
M.A.C. Perryman ed., ESA, in press 
\bibitem[]{} Freedman, W.L., \& Madore, B.F., 1990, ApJ, 365, 186
\bibitem[]{} Fusi Pecci, F. et al. 1996, AJ, 112, 1461
\bibitem[]{} Gratton, R. G., Carretta, E., \& Castelli, F. 1997a, A\&A, 314, 191
\bibitem[]{} Gratton, R. G., Fusi Pecci, F., Carretta, E., Clementini, G.,
Corsi, C. E., \& Lattanzi, M. 1997b, ApJ, in press (December 20th issue)
\bibitem[]{} Gratton, R. G., Clementini, G., Fusi Pecci, F., Carretta, E., 
1997c, in Views on Distance Indicators, F. Caputo ed., Mem. S.A.It., in press
\bibitem[]{} Gould, A., \& Uza, O. 1997, astro-ph/9705051
\bibitem[]{} Gray, R. O., Corbally, C. J., \& Philip, A. G. D. 1996, AJ, 112,
2291 
\bibitem[]{} Hauck, B., \& Mermilliod, M. 1990, A\&AS, 86, 107 
\bibitem[]{} Kinman, T. D., Stryker, L. L., Hesser, J. E., Graham, J. A., 
Walker, A. R., Hazen, M. L., \& Nemec, J. M. 1991, PASP, 103, 1279 
\bibitem[]{} Kraft, R.P. 1994, PASP, 106, 553
\bibitem[]{} Layden, A. C. 1994, AJ, 108, 1016
\bibitem[]{} Layden, A. C., Hanson, R. B., Hawley, S. L., Klemola, A. R., \&
Hanley, C. J. 1996, AJ, 112, 2110 
\bibitem[]{} Liu, T., \& Janes, K. A. 1990, ApJ, 360, 561
\bibitem[]{} Lutz, T.E., \& Kelker, D.H. 1973, PASP, 85, 573
\bibitem[]{} McNamara, D.H. 1997, PASP, 109, 857
\bibitem[]{} Olsen, E. H. 1993, A\&AS, 102, 89
\bibitem[]{} Olsen, E. H. 1994a, A\&AS, 104, 429
\bibitem[]{} Olsen, E. H. 1994b, A\&AS, 106, 257
\bibitem[]{} Panagia, N., Gilmozzi, R., \& Kirshner, R. P. 1997, in SN1987a:
Ten Years After, eds. M. Phillips \& N. Suntzeff, ASP Conf. Ser., in press 
\bibitem[]{} Perryman, M. A. C. et al. 1997, A\&A, 323, L49
\bibitem[]{} Pont et al. 1997, A\&A, in press
\bibitem[]{} Preston, G. W., Schectman, S. A., \& Beers, T. C. 1991, ApJ, 375,
121 
\bibitem[]{} Reid, I. N. 1997, AJ, 114, 161
\bibitem[]{} Rich, D. M., Mighell, K. J., \& Neill, J. D. 1996, in Formation of
the Galactic Halo... Inside and Out, H. Morrison \& A. Sarajedini eds., ASP 
Conf. Ser. 92, p. 544
\bibitem[]{} Sandage, A.R. 1993, AJ, 106, 703
\bibitem[]{} Sandquist, E. L., Bolte, M., Stetson, P. B., \& Hesser, J. E. 1996,
ApJ, 470, 910 
\bibitem[]{} Sneden, C., Kraft, R.P., Prosser, C.F., \& Langer, G.E. 1992, AJ,
104, 2121
\bibitem[]{} Stetson, P. B. 1991, AJ, 102, 589
\bibitem[]{} Storm, J., Carney, B. W., \& Latham, D. W. 1994, A\&A, 290, 443
\bibitem[]{} Sweigart, A. V. 1997, in The Third Conference on Faint Blue Stars,
ed. A. G. D. Philip, astro-ph/9708164
\bibitem[]{} van Leeuwen, F., Feast, M. W., Whitelock, P. A., \& Yudin 1997,
MNRAS, 287, p55 
\bibitem[]{} Walker, A. R. 1992, ApJ, 390, L81

\end{thebibliography}
\end{document}